\documentstyle[12pt]{article}
\newcommand{\bd}{\begin{document}}
\newcommand{\ed}{\end{document}}
\newcommand{\bc}{\begin{center}}
\newcommand{\ec}{\end{center}}
\newcommand{\be}{\begin{eqnarray}}
\newcommand{\ee}{\end{eqnarray}}
\newcommand{\text}{\rm }
\newcommand{\func}{\rm }
\renewcommand{\thefootnote}{\alph{footnote}}
\newcommand{\se}{\section}
\newcommand{\sse}{\subsection}
\newcommand{\bi}{\bibitem}
\def\figcap{\section*{Figure Captions\markboth
     {FIGURECAPTIONS}{FIGURECAPTIONS}}\list
     {Figure \arabic{enumi}:\hfill}{\settowidth\labelwidth{Figure 999:}
     \leftmargin\labelwidth
     \advance\leftmargin\labelsep\usecounter{enumi}}}
\let\endfigcap\endlist \relax

\begin{document}

\begin{titlepage}

 \vskip 0.5in
 \null
\begin{center}
 \vspace{.15in}
{\LARGE {\bf
Lepton Asymmetries in Heavy Baryon Decays of $\Lambda_b\to\Lambda l^+l^-$}
}\\
\vspace{1.0cm}  \par
 \vskip 2.1em
 {\large
  \begin{tabular}[t]{c}
{\bf Chuan-Hung Chen$^a$ and C.~Q.~Geng$^b$}
\\
\\
       {\sl ${}^a$Department of Physics, National Cheng Kung University}
\\   {\sl  $\ $Tainan, Taiwan,  Republic of China }
\\
\\
{\sl ${}^b$Department of Physics, National Tsing Hua University}
\\  {\sl  $\ $ Hsinchu, Taiwan, Republic of China }
\\
   \end{tabular}}
 \par \vskip 5.3em

\date{\today}
 {\Large\bf Abstract}
\end{center}

We study the dilepton forward-backward and the longitudinal,
normal and transverse lepton polarization asymmetries in the
heavy baryon decays of $\Lambda_b\to\Lambda l^+l^-$. We show that
the asymmetries have a less dependence on the non-perturbative
QCD effects. In the standard model, we find that the integrated
forward-backward asymmetries (FBAs) and three components of the
polarizations in the QCD sum rule approach (pole model) are
$-0.13\ (-0.12)$ and $(58.3,-9.4,-0.07)\%$
($(58.3,-12.6,-0.07)\%$) for $\Lambda_b\to\Lambda\mu^+\mu^-$ and
$-0.04\ (-0.03)$ and $(10.9,-10.0,-0.39)\%$
($(10.9,-0.2,-0.34)\%$) for $\Lambda_b\to\Lambda\tau^+\tau^-$,
respectively.

\end{titlepage}

\section{ Introduction}

It is known that the FBAs of the dileptons in the inclusive decays of $b\to
sl^+l^-$ provide us with information on the short-distance (SD)
contributions, which are dominated by the top quark loops in the standard
model \cite{AMM}. The longitudinal lepton polarizations in $b\to sl^+l^-$,
which are another parity violating observables, are also interesting
asymmetries. In particular, the tau polarization in $b\to s\tau^-\tau^-$
could be accessible to the B-Factories \cite{JLH,Sehgal}. It is noted that
the FBAs of the exclusive decays $B\to M l^+l^-$ are identically zero when $%
M $ are pseudoscalar mesons such as $\pi$ and $K$ but nonzero for $M$ being
vector mesons such as $\rho$ and $K^*$. However, the longitudinal lepton
polarizations \cite{geng1} as well as other components \cite{geng2} are
nonzero for both types of the exclusive B meson decay modes.

In this paper, we study the dilepton forward-backward and various lepton
polarization asymmetries in the heavy baryon decays of $\Lambda_b\to\Lambda
l^+l^-$. To study these baryonic decays, one of the most difficulties is to
evaluate the hadronic matrix elements. It is known that there are many form
factors for the $\Lambda_b\to \Lambda$ transition, which are hard to be
calculated since they are related to the non-perturbative effect of QCD.
However, in heavy particle decays, the heavy quark effective theory (HQET)
could reduce the number of form factors and supply the information with
respect to their relative size \cite{Huang,chen1,chen2}. With the HQET, we
shall use the QCD sum rule approach \cite{Huang} and the pole model \cite{MR}
in our numerical calculations for the form factors.

The paper is organized as follows. In Sec.~2, we study the effective
Hamiltonian for the decays of $\Lambda_b \to \Lambda l^+ l^- \
(l=e,\mu,\tau) $ and form factors in the $\Lambda_b \to \Lambda$ transition.
In Sec.~3, we derive the general forms of the lepton polarization and
dilepton forward-backward asymmetries in $\Lambda_b\to\Lambda l^+ l^-$. We
give our numerical analysis in Sec.~4. In Sec.~5, we present our conclusions.

\section{Effective Hamiltonian and Form factors}

To study the heavy baryon decays of $\Lambda_b\to\Lambda l^+l^-$ ($l=e$ or $%
\mu$ or $\tau$), we start with the effective Hamiltonian for the $b$-quark
decay of $b\to s l^{+}l^{-}$, given by

\begin{equation}
{\cal H}=-4\frac{G_{F}}{\sqrt{2}}V_{tb}V_{ts}^{*}\sum_{i=1}^{10}C_{i}\left(
\mu \right) O_{i}\left( \mu \right) \,  \label{Ham}
\end{equation}
where $G_{F}$ is the Fermi constant, $V_{ij}$ are the CKM matrix elements,
and $C_{i}(\mu )$ and $O_{i}(\mu )$ are the expressions for the renormalized
Wilson coefficients and operators, whose expressions can be found in Ref.
\cite{Buras}, respectively. In terms of the Hamiltonian in Eq. (\ref{Ham}),
the free quark decay amplitude is written as
\begin{eqnarray}
{\cal M}\left( b\rightarrow sl^{+}l^{-}\right) &=&\frac{G_{F}\alpha _{em}}{%
\sqrt{2}\pi }V_{tb}V_{ts}^{*}\left[ \bar{s}\left( C_{9}^{eff}\left( \mu
\right) \gamma _{\mu }P_{L}-\frac{2m_{b}}{q^{2}}C_{7}\left( \mu \right)
i\sigma _{\mu \nu }q^{\nu }P_{R}\right) b\;\bar{l}\gamma ^{\mu }l\right.
\nonumber \\
&&\left. +\bar{s}C_{10}\gamma _{\mu }P_{L}b\;\bar{l}\gamma ^{\mu }\gamma
_{5}l\right]  \label{Am}
\end{eqnarray}
with $P_{L(R)}=(1\mp \gamma _{5})/2$. We note that in Eq. (\ref{Am}), only
the term associated with the Wilson coefficient $C_{10}$ is independent of
the $\mu $ scale. We also note that the dominant contribution to the decay
rate is from the long-distance (LD), such as that from the $c\bar{c}$
resonant states of $\Psi ,\Psi ^{\prime }...etc$. It is known that to find
out the LD effects for the B-meson decays, in the literature \cite
{AMM,Sehgal,geng1,DTP,LMS,OT}, both the factorization assumption (FA) and
the vector meson dominance (VMD) approximation have been used. For the LD
contributions in baryonic decays, we assume that the parametrization is the
same as that in the B meson decays. Hence, we may include the resonant
effect (RE) by absorbing it to the corresponding Wilson coefficients. The
effective Wilson coefficient of $C_{9}^{eff}$ has the standard form
\[
C_{9}^{eff}=C_{9}\left( \mu \right) +\left( 3C_{1}\left( \mu \right)
+C_{2}\left( \mu \right) \right) \left( h\left( x,s\right) +\frac{3}{\alpha
_{em}^{2}}\sum_{j=\Psi ,\Psi ^{\prime }}k_{j}\frac{\pi \Gamma \left(
j\rightarrow l^{+}l^{-}\right) M_{j}}{q^{2}-M_{j}^{2}+iM_{j}\Gamma _{j}}%
\right)
\]
where $h(x,s)$ describes the one-loop matrix elements of operators $O_{1}=%
\bar{s}_{\alpha }\gamma ^{\mu }P_{L}b_{\beta }\ \bar{c}_{\beta }\gamma _{\mu
}P_{L}c_{\alpha }$ and $O_{2}=\bar{s}\gamma ^{\mu }P_{L}b\ \bar{c}\gamma
_{\mu }P_{L}c$ as shown in Ref. \cite{Buras}, $M_{j}\ (\Gamma _{j})$ are the
masses (widths) of intermediate states, and the factors $k_{j}$ are
phenomenological parameters for compensating the approximations of FA and
VMD and reproducing the correct branching ratios of $B(B\to J/\Psi X\to
l^{+}l^{-}X)=B(B\to J/\Psi X)\times B(J/\Psi \to l^{+}l^{-})$. In this paper
we take the Wilson coefficients at the scale of $\mu \sim m_{b}\sim 5.0$ GeV
and their values are taking to be $C_{1}\left( m_{b}\right) =-0.226,$ $%
C_{2}\left( m_{b}\right) =1.096,$ $C_{7}\left( m_{b}\right) =-0.305,$ $%
C_{9}\left( m_{b}\right) =4.186,$ and $C_{10}\left( m_{b}\right) =-4.599$,
respectively.

It is clear that one of the main theoretical uncertainties in studying
exclusive decays arises from the calculation of form factors.
%In general there are many form factors in exclusive baryon decays.
%However, the number of the form factors can be reduced by the heavy quark
%effective theory (HQET).
With the HQET, the hadronic matrix elements for the heavy baryon decays
could be parametrized as follows \cite{MR}
\begin{equation}
<\Lambda (p,s)\ |\ \bar{s}\ \Gamma \ b\ |\ \Lambda _{b}(v,s^{\prime })>=\bar{%
u}_{\Lambda }(p,s)\ \left\{ F_{1}(q^{2})+\not{v}\ F_{2}(q^{2})\right\} \
\Gamma \ u_{\Lambda _{b}}(v,s^{\prime })  \label{hq}
\end{equation}
where $v=p_{\Lambda _{b}}/M_{\Lambda _{b}}$ is the four-velocity of the
heavy baryon, $q^2=(p_{\Lambda_b}-p_{\Lambda})^2$ is the square of the
momentum transform, and $\Gamma $ denotes the possible Dirac matrix.
% we have used the particle name to denote its spinor state.
Note that in terms of the HQET there are only two independent form factors, $%
F_1$ and $F_2$, in Eq. (\ref{hq}) for each $\Gamma $. In the following, we
shall use $F_1$ and $R\equiv F_2/F_1$ as the two independent parameters and
adopt the HQET approximation to analyze the behavior of $\Lambda _{b}\to
\Lambda l^{+}l^{-}$.

 From Eqs. (\ref{Am}) and (\ref{hq}), the transition matrix element for $%
\Lambda _{b}\left( p_{\Lambda _{b}}\right) \rightarrow \Lambda \left(
p_{\Lambda }\right) l^{+}\left( p_{+}\right) l^{-}\left( p_{-}\right) $ can
be expressed as

\begin{equation}
{\cal M}\left( \Lambda _{b}\rightarrow \Lambda l^{+}l^{-}\right) =\frac{%
G_{F}\alpha _{em}}{\sqrt{2}\pi }V_{tb}V_{ts}^{*}\left[ H_{1\mu }L_{V}^{\mu
}+H_{2\mu }L_{A}^{\mu }\right]  \label{hamp}
\end{equation}
with
\begin{eqnarray}
L_{V} &=&\bar{l}\gamma ^{\mu }l\,,  \nonumber \\
L_{A} &=&\bar{l}\gamma ^{\mu }\gamma _{5}l\,,  \nonumber \\
H_{1\mu } &=&\bar{\Lambda}\gamma _{\mu }\left( A_{1}P_{R}+B_{1}P_{L}\right)
\Lambda _{b}+\bar{\Lambda}i\sigma _{\mu \nu }q^{\nu }\left(
A_{2}P_{R}+B_{2}P_{L}\right) \Lambda _{b},  \nonumber \\
H_{2\mu } &=&E_{1}\bar{\Lambda}\gamma _{\mu }P_{L}\Lambda _{b}+E_{2}\bar{%
\Lambda}i\sigma _{\mu \nu }q^{\nu }P_{L}\Lambda _{b}+E_{3}q_{\mu }\bar{%
\Lambda}( P_{L} \Lambda _{b}
\end{eqnarray}
where one has
\begin{eqnarray}
q &=&p_{\Lambda _{b}}-p_{\Lambda }=p_{+}+p_{-}\,,  \nonumber \\
A_{i} &=&-\frac{2m_{b}}{q^{2}}C_{7}f_{i}^{T}{},\ B_{i}=C_{9}^{eff}f_{i},\
E_{i}=C_{10}f_{i}\,,  \nonumber \\
f_{1} &=&f_{2}^{T}=F_{1}+\sqrt{r}RF_{1},\ f_{2}=f_{3}=\frac{RF_{1}}{%
M_{\Lambda _{b}}}.
\end{eqnarray}
%Note that $A_{2}$ and $B_{2}$ have no contributions to the dileptonic
%decay due to $q_{\mu }L_{V}^{\mu }=0$.

\section{Lepton Asymmetries}

In this section we present the formulas for the forward-backward and the
longitudinal, normal and transverse lepton polarization asymmetries in $%
\Lambda _{b}(p_{\Lambda _{b}})\to \Lambda (p_{\Lambda
})l^{+}(p_{+},s_{+})l^{-}(p_{-})$. We shall concentrate on the $l^{+}$ spin
for the polarizations. To do this, we write the $l^{+}$ four-spin vector in
terms of a unit vector, $\hat{\xi}$, along the $l^{+}$ spin in its rest
frame, as
\begin{eqnarray}
s_{+}^{0}\,=\,\frac{\vec{p}_{+}\cdot \hat{\xi}}{m_{l}},\qquad \vec{s}%
_{+}\,=\,\hat{\xi}+\frac{s_{+}^{0}\,}{E_{l^{+}}+m_{l}}\vec{p}_{+},
\end{eqnarray}
and choose the unit vectors along the longitudinal, normal, and transverse
components of the $l^{+}$ polarization to be
\begin{eqnarray}
\hat{e}_{L} &=&\frac{\vec{p}_{+}}{\left| \vec{p}_{+}\right| },  \nonumber \\
\hat{e}_{N} &=&\frac{\vec{p}_{+}\times \left( \vec{p}_{\Lambda }\times \vec{p%
}_{+}\right) }{\left| \vec{p}_{+}\times \left( \vec{p}_{\Lambda }\times \vec{%
p}_{+}\right) \right| },  \nonumber \\
\hat{e}_{T} &=&\frac{\vec{p}_{\Lambda }\times \vec{p}_{+}}{\left| \vec{p}%
_{\Lambda }\times \vec{p}_{+}\right| }\,,  \label{uv}
\end{eqnarray}
respectively. The partial decay width for $\Lambda _{b}\rightarrow \Lambda \
l^{+}\ l^{-}$ is given by
\begin{eqnarray}
d\Gamma &=&\frac{1}{4M_{\Lambda _{b}}}\left| {\cal M}\right| ^{2}\left( 2\pi
\right) ^{4}\delta \left( p_{\Lambda _{b}}-p_{\Lambda
}-p_{l^{+}}-p_{l^{-}}\right)  \nonumber \\
&&\times \frac{d\vec{p}_{\Lambda }}{\left( 2\pi \right) ^{3}2E_{\Lambda }}%
\frac{d\vec{p}_{l^{+}}}{\left( 2\pi \right) ^{3}2E_{1}}\frac{d\vec{p}_{l^{-}}%
}{\left( 2\pi \right) ^{3}2E_{2}}  \label{Dr0}
\end{eqnarray}
with
\begin{equation}
|{\cal M}|^{2}=\frac{1}{2}\left| {\cal M}^{0}\right| ^{2}\left[ 1+\left( P_L%
\hat{e}_{L}+P_N\hat{e}_{N}+ P_T\hat{e}_{T}\right) \cdot \hat{\xi}\right] \,,
\label{M1}
\end{equation}
where $|{\cal M}^{0}|^{2}$ is related to the decay rate for the unpolarized $%
l^{+}$ and $P_i\ (i=L,N,T)$ are the longitudinal, normal and transverse
polarizations of $l^{+}$, respectively. Introducing dimensionless variables
of $\lambda _{t}=V_{tb}V_{ts}^{*}$, $r=M_{\Lambda }^{2}/M_{\Lambda _{b}}^{2}$%
, $\hat{m}_{l}=m_{l}/M_{\Lambda _{b}}$, $\hat{m}_{b}=m_{b}/M_{\Lambda _{b}}$%
, $\hat{s}=q^{2}/M_{\Lambda _{b}}^{2}$ and $\hat{t}=p_{\Lambda _{b}}\cdot
p_{\Lambda }/M_{\Lambda _{b}}^{2}=(1+r-\hat{s})/2$, using the transition
matrix element of Eq. (\ref{hamp}), and integrating the angle dependence of
the lepton, the differential decay width in Eq. (\ref{Dr0}) becomes
\begin{equation}
d\Gamma =\frac{1}{2}d\Gamma ^{0}\left[ 1+\vec{P}\cdot \vec{\xi}\right]
\label{diffrate}
\end{equation}
with
\begin{eqnarray}
d\Gamma^0 &=&\frac{G_{F}^{2}\alpha _{em}^{2}\lambda _{t}^{2}}{384\pi ^{5}}%
M_{\Lambda _{b}}^{5}\sqrt{\phi \left( \hat{s}\right) } \sqrt{1-\frac{4\hat{m}%
_l^2}{\hat{s}}}R_{\Lambda _{b}}\left( \hat{s}\right)\, d\hat{s}  \label{rate}
\end{eqnarray}
and
\begin{eqnarray}
\vec{P} &=& P_{L}\hat{e}_{L}+P_{N}\hat{e}_{N}+P_{T}\hat{e}_{T},
\end{eqnarray}
where
\begin{eqnarray}
\phi \left( \hat{s}\right) =(1-r)^{2}-2\hat{s}(1+r)+\hat{s}^{2}
\end{eqnarray}
and
\begin{eqnarray}
R_{\Lambda_b}(\hat{s}) &=& 4\frac{\hat{m}_{b}^{2}}{\widehat{s}}%
|C_{7}|^{2}F_{1}^{2}\left\{ -\left( 1-R^{2}\right) \left[ \hat{s}\ \hat{t}%
-4\left( 1-\hat{t}\right) \left( \hat{t}-r\right) \right] \right.  \nonumber
\\
&& -2R\left( \sqrt{r}+R\hat{t}\right) \left( \hat{s}\ -4(1-\hat{t}%
)^{2}\right) +8\frac{\hat{m}_{l}^{2}}{\hat{s}}\left[ \left( 1-R^{2}\right)
\left( 1-\hat{t}\right) \left( \hat{t}-r\right)\right.  \nonumber \\
&& \left.\left. +2R\left( \sqrt{r}+R\hat{t}\right) \left( 1-\hat{t}\right)
^{2}\right] -2\hat{m}_{l}^{2}\left( \left( 1+R^{2}\right) \ \hat{t}+2R\sqrt{r%
}\right) \right\}  \nonumber \\
&&+12\hat{m}_{b}{Re}C_{9}^{eff}C_{7}^{*}\left( 1+2\frac{\hat{m}_{l}^{2}}{%
\hat{s}}\right) F_{1}^{2}\left[ \left( 1-R^{2}\right) \left( \hat{t}%
-r\right) +2R\left( \sqrt{r}+R\hat{t}\right) \left( 1-\hat{t}\right) \right]
\nonumber \\
&&+\left( |C_{9}^{eff}|^{2}+|C_{10}|^{2}\right) F_{1}^{2}\left\{ \left( 1-4%
\frac{\hat{m}_{l}^{2}}{\hat{s}}\right) \left[ \left( 1+R^{2}\right) \ \hat{t}%
+2R\sqrt{r}\right] \right.  \nonumber \\
&&\left. +2\left( 1+2\frac{\hat{m}_{l}^{2}}{\hat{s}}\right) \left( 1-\hat{t}%
\right) \left[ \left( \hat{t}-r\right) \left( 1-R^{2}\right) +2R\left( \sqrt{%
r}+R\hat{t}\right) \left( 1-\hat{t}\right) \right] \right\} ,  \nonumber \\
&&+6\hat{m}_{l}^{2}\left( |C_{9}^{eff}|^{2}-|C_{10}|^{2}\right)
F_{1}^{2}\left[ \left( 1+R^{2}\right) \ \hat{t}+2R\sqrt{r}\right]\,.
\label{RLambdab}
\end{eqnarray}
In Eqs. (\ref{rate}) and (\ref{RLambdab}), the allowed range of $\hat{s}$ is
\begin{eqnarray}
4\hat{m}_{l}^{2}\leq \hat{s}\leq \left( 1-\sqrt{r}\right) ^{2}\,.
\end{eqnarray}
%and $R=F_2(q^2)/F_1(q^2)$.

Defining the longitudinal, normal and transverse $l^{+}$ polarization
asymmetries by
\begin{equation}
P_{i}\left( \hat{s}\right) =\frac{d\Gamma \left( \hat{e}_{i}\cdot \hat{\xi}%
=1\right) -d\Gamma \left( \hat{e}_{i}\cdot \hat{\xi}=-1\right) }{d\Gamma
\left( \hat{e}_{i}\cdot \hat{\xi}=1\right) +d\Gamma \left( \hat{e}_{i}\cdot
\hat{\xi}=-1\right) }\,,  \label{pasy}
\end{equation}
from Eq. (\ref{diffrate}) we find that
\begin{eqnarray}
P_{L}\left( \hat{s}\right) &=&-\sqrt{1-\frac{4\hat{m}_{l}^{2}}{\hat{s}}}%
\frac{R_{L}\left( \hat{s}\right) }{R_{\Lambda _{b}}\left( \hat{s}\right) },
\label{pl} \\
P_{N}\left( \hat{s}\right) &=&\frac{3}{4}\pi \hat{m}_{l}\sqrt{\frac{\phi
\left( \hat{s}\right) }{\hat{s}}}\frac{R_{N}\left( \hat{s}\right) }{%
R_{\Lambda _{b}}\left( \hat{s}\right) },  \label{pn} \\
P_{T}\left( \hat{s}\right) &=&\frac{3}{4}\pi \hat{m}_{l}\sqrt{\hat{s}\ \phi
\left( \hat{s}\right) }\sqrt{1-\frac{4\hat{m}_{l}^{2}}{\hat{s}}}\frac{%
R_{T}\left( \hat{s}\right) }{R_{\Lambda _{b}}\left( \hat{s}\right) }\,,
\label{pt}
\end{eqnarray}
where
\begin{eqnarray}
R_{L}\left( \hat{s}\right) &=&F_{1}^{2}{Re}C_{9}^{eff}C_{10}^{*}\left[
\left( 1-R^{2}\right) \left( \left( 1-r\right) ^{2}+\hat{s}\left( 1+r\right)
-2\,\hat{s}^{2}\right) \right.  \nonumber \\
&&\left. +2R\left( \sqrt{r}+R\hat{t}\right) \left( 2\,\hat{s}+\left( 1-r+%
\hat{s}\right) ^{2}\right) \right]  \nonumber \\
&&+6F_{1}^{2}{Re}C_{10}C_{7}^{*}\hat{m}_{b}\left[ \left( 1-r-\hat{s}\right)
(1-R^{2})+2R(\sqrt{r}+R\hat{t})\left( 1-r+\hat{s}\right) \right] ,  \nonumber
\\
R_{N}\left( \hat{s}\right) &=&4F_{1}^{2}\frac{\hat{m}_{b}^{2}}{\hat{s}}%
|C_{7}|^{2}\left[ \left( 1-R^{2}\right) \left( 1-r\right) +2R(\sqrt{r}+R\hat{%
t})\left( 1-r+s\right) \right]  \nonumber \\
&&+F_{1}^{2}\left( 1-R^{2}\right) |C_{9}^{eff}|^{2}\ \hat{s}+F_{1}^{2}{Re}%
C_{9}^{eff}C_{10}^{*}\left[ \left( 1-r\right) \left( 1-R^{2}\right) \right.
\nonumber \\
&&\left. +2\left( 1-r+\hat{s}\right) R(\sqrt{r}+R\hat{t})\right]  \nonumber
\\
&&+2F_{1}^{2}\hat{m}_{b}\left( 2{Re}C_{9}^{eff}C_{7}^{*}+{Re}%
C_{10}C_{7}^{*}\right) \left( 1-R^{2}+2R(\sqrt{r}+R\hat{t})\right) ,
\nonumber \\
R_{T}\left( \hat{s}\right) &=&F_{1}^{2}\frac{2\hat{m}_{b}}{\hat{s}}{Im}%
C_{7}C_{10}^{*}\left( 1-R^{2}+2R\left( \sqrt{r}+R\hat{t}\right) \right)
+F_{1}^{2}{Im}C_{9}^{eff}C_{10}^{*}\left( 1-R^{2}\right) .  \label{Rlnt}
\end{eqnarray}
%\end{eqnarray*}
We note that the transverse part of the lepton polarization in Eq. (\ref{pt}%
) is a T-odd observable.

The differential and normalized dilepton forward-backward asymmetries (FBAs)
for the decay of $\Lambda_b\to\Lambda l^+l^-$ as a function of $\hat{s}$ are
defined by
\begin{eqnarray}
\frac{dA_{FB}\left( \hat{s}\right) }{d\hat{s}} &=&\left[ \int_{0}^{1}d\cos
\theta \ \frac{d^{2}\Gamma \left( \hat{s}\right) }{d\hat{s}d\cos \theta }%
-\int_{-1}^{0}d\cos \theta \ \frac{d^{2}\Gamma \left( \hat{s}\right) }{d\hat{%
s}d\cos \theta }\right]  \label{diffba1}
\end{eqnarray}
and
\begin{eqnarray}
{\cal A}_{FB}\left( \hat{s}\right) &=&\frac{1}{d\Gamma \left( \hat{s}\right)
/d\hat{s}}\left[ \int_{0}^{1}d\cos \theta \ \frac{d^{2}\Gamma \left( \hat{s}%
\right) }{d\hat{s}d\cos \theta }-\int_{-1}^{0}d\cos \theta \ \frac{%
d^{2}\Gamma \left( \hat{s}\right) }{d\hat{s}d\cos \theta }\right] \,,
\label{fba1}
\end{eqnarray}
respectively, where $\theta $ is the angle of $l^{+}$ with respect to $%
\Lambda_b$ in the rest frame of the lepton pair. Explicitly, we obtain
\begin{eqnarray}
\frac{dA_{FB}\left( \hat{s}\right) }{d\hat{s}} &=&\frac{G_{F}^{2}\alpha
_{em}^{2}\lambda _{t}^{2}}{2^{8}\pi ^{5}}M_{\Lambda _{b}}^{5}\phi \left(
\hat{s}\right) \left( 1-4\frac{\hat{m}_{l}^{2}}{\hat{s}}\right) R_{FB}\left(
\hat{s}\right)  \label{diffba2}
\end{eqnarray}
and
\begin{eqnarray}
{\cal A}_{FB}\left( \hat{s}\right) &=& \frac{3}{2}\sqrt{\phi \left( \hat{s}%
\right) }\sqrt{1-\frac{4\hat{m}_{l}^{2}}{s}}\frac{R_{FB}\left( \hat{s}%
\right) }{R_{\Lambda _{b}}\left( \hat{s}\right) }  \label{fba2}
\end{eqnarray}
where
\begin{eqnarray}
R_{FB}\left( \hat{s}\right) &=&F_{1}^{2}\left( 1-R^{2}\right) \left[ 2\hat{m}%
_{b}{Re}C_{10}C_{7}^{*}\left( 1+2\frac{R\sqrt{r}+R^{2}\hat{t}}{1-R^{2}}%
\right) +\hat{s}{Re}C_{9}^{eff}C_{10}^{*}\right]\,.  \label{rfba}
\end{eqnarray}
 From Eqs. (\ref{RLambdab}), (\ref{pl})-(\ref{Rlnt}), and
(\ref{fba2})-(\ref
{rfba}), we see that $P_i\ (i=L,N,T)$ and ${\cal A}_{FB}$ depend only on $R$
since the factor $F_1^2$ is canceled out. Thus, once one gets the value of $%
R $, the only uncertainty for the asymmetries is from the Wilson
coefficients. It is interesting to note that these asymmetries are sensitive
to the chiral structure of electroweak interactions since they are related
to the products of $C_9C_7^*$, $C_{10}C_{7}^{*}$ and $C_{9}C_{10}^{*}$.

\section{Numerical Analysis}

In our numerical calculations, the Wilson coefficients are evaluated at the
scale $\mu \simeq m_{b}$ and the other parameters are listed in Table 1 of
Ref. \cite{chen2}. For the form factors in the $\Lambda _{b}\rightarrow
\Lambda $ transition, we use the results from both the QCD sum rule approach
\cite{Huang} and the pole model \cite{MR}. In the QCD sum rule approach we
use the form
\begin{equation}
F_{i}(q^{2})=\frac{F_{i}(0)}{1+aq^{2}+bq^{4}}\,,  \label{qcdff}
\end{equation}
where the parameters in Eq. (\ref{qcdff}) are shown in Table 1. From the
Table, we find that $R(0)=F_{2}(0)/F_{1}(0)=-0.17$ and $R(q_{\max
}^{2})=-0.44$
 which are consistent with the CLEO result of
 $R=-0.25\pm 0.14\pm 0.08$ \cite{CLEO}.
\begin{table}[h]
\caption{ Form Factors in the QCD sum rule approach.}
\begin{center}
\begin{tabular}{|c|c|c|}
\hline
& $F_{1}$ & $F_{2}$ \\ \hline
$q^2=0$ & $0.462$ & $-0.077$ \\ \hline\hline
$a$ & $-0.0182$ & $-0.0685$ \\ \hline
$b$ & $-0.000176$ & $0.00146$ \\ \hline
\end{tabular}
\end{center}
\end{table}
In the pole model, we adopt
\be
F_{i}(q^{2})=N_{i}\left( \frac{\Lambda _{QCD}}{\Lambda _{QCD}+z}\right) ^{2}
\ee
where $z=p_{\Lambda }\cdot p_{\Lambda _{b}}/M_{\Lambda
_{b}}=(1+r-q^{2}/M_{\Lambda _{b}}^{2})M_{\Lambda _{b}}/2$ and
 $\Lambda_{QCD}$ is chosen around $200\ MeV$.
Assuming the form factors for the transition of
 $\Lambda _{c}\rightarrow \Lambda $ are similar to that of
$\Lambda_{b}\rightarrow \Lambda$  and using $R=-0.25$ \cite{CLEO}
and the branching ratio of $\Lambda _{c}^{+}\rightarrow \Lambda e^{+}\nu
_{e}$, we obtain that $N_{1,2}$ are $(52.32,-13.08)$
\cite{chen2}.

\subsection{Forward-backward Asymmetries}

 From Eqs. (\ref{diffba2}) and (\ref{fba2}), we see that the FBAs for the
light charged lepton modes of $\Lambda _{b}\rightarrow \Lambda l^{+}l^{-}$ ($%
l=e$ and $\mu $) are close to each other. As a result, we shall not mention
the electron mode of $\Lambda _{b}\to \Lambda e^{+}e^{-}$. In Figures 1 and
2, we show ${\cal A}_{FB}(\Lambda _{b}\to \Lambda l^{+}l^{-})$ as a function
of dimensionless variable $\hat{s}$ for $l=\mu $ and $\tau $, respectively.
 From Figure 1(a), we see that ${\cal A}_{FB}(\Lambda _{b}\to \Lambda \mu
^{+}\mu ^{-})$ has a zero value at $\hat{s}_{0}$ which satisfies the
condition
\be
{Re}C_{9}^{eff}C_{10}^{*}&=&-\frac{2\hat{m}_{b}}{\hat{s}_{0}}{Re}%
C_{7}C_{10}^{*}\frac{1-R^{2}+2R\left( \sqrt{r}+R\ \hat{t}\right) }{\left(
1-R^{2}\right) }.
\label{rec9c10}
\ee
Furthermore, we find
%If we use $R\sim -0.25$ and $\sqrt{r}\sim 0.20$,
 that the contributions from the pole and QCD sum rule models to
FBAs overlap at the low $q^{2}$ region so that in both models Eq.
(\ref{rec9c10}) can be simplified to
\be
{Re}C_{9}^{eff}C_{10}^{*}&\simeq& -\frac{2\hat{m}_{b}}{\hat{s}_{0}}{Re}%
C_{7}C_{10}^{*}\,,
\label{mc9c10}
\ee
which is independent of the hadronic form factors.
Explicitly, from Figure 1(a),
 in the standard model
we get that $\hat{s}_{0}$ is $0.109$ and $0.114$ with and without $R$
terms for excluding LD effects, and  $0.098$ and $0.102$ for including
LD effects, respectively.
 It is clear that the zero point of
${\cal A}_{FB}(\Lambda _{b}\to \Lambda \mu ^{+}\mu ^{-})$ is
mainly affected  by the weak Wilson coefficients of $C_{7}$ and $C_{9}$
that are sensitive to physics beyond the standard model.
For example, if one of $C_{7}$ and $C_{9}$ has an opposite sign to that in
the standard model, the condition for the zero point in Eq.
(\ref{mc9c10}) will not be satisfied. Therefore, measuring a
sizable value of the FBA around $\hat{s}_{0}$ is a clear
indication of new physics. This result is similar to
$B\rightarrow K^{*}l^{+}l^{-}$ decays mentioned by \cite{Ali}
with large energy effective theory (LEET) \cite {Charles}. We
 note that the vanishing of the FBAs in the inclusive decays of
$b\to (s,d)l^{+}l^{-}$ and the exclusive ones of $B\to
(K^{*},\rho )l^{+}l^{-}$ were first studied by Burdman
\cite{Burdman}. Our conclusion for the baryonic decays coincides
with that in Ref. \cite{Burdman}.

 From the figures, we find that there is no much difference for the FBAs
between the QCD sum rule approach and the pole model at the lower values of $%
q^{2}$, especially for that in the muon mode. By taking $R$ to be zero, the
distributions for both models in Figures 1 and 2 should be identical. Thus,
the differences for the FBAs in the different QCD models actually reflect
the effects of the ratio $R$. The insensitivity to the form factors for the
FBAs provides us a candidate to test the standard model.

In Figure 3, we show the differential FBA of $dA_{FB}(\hat{s})/d\hat{s}$
which, unlike ${\cal A}_{FB}$, is insensitive to R. This can be understood
that due to Eqs. (\ref{diffba2}) and (\ref{fba2}) it is proportional to $%
R_{FB}(\hat{s})$ in which the terms with $F_1^2$ are the dominant
contributions and those with $R$ are negligible since these terms are
related to either $R^2$ or $R\sqrt{r}$, which are small.

We now define the integrated FBA to be
\begin{eqnarray}
\overline{{\cal A}}_{FB}=\int_{4\hat{m}_{l}^{2}}^{\hat{s}_{\max }}d\hat{s}%
{\cal A}_{FB}\left( \hat{s}\right)
\end{eqnarray}
where $\hat{s}_{\max }=\left( 1-\sqrt{r}\right) ^{2}$. Without LD
contributions, in the standard model we find that
\begin{eqnarray}
\overline{{\cal A}}_{FB}(\Lambda _{b}\to \Lambda \mu ^{+}\mu ^{-})=-0.13\
(-0.12)
\end{eqnarray}
and
\begin{eqnarray}
\overline{{\cal A}}_{FB}(\Lambda _{b}\to \Lambda \tau ^{+}\tau ^{-})=-0.04\
(-0.03)
\end{eqnarray}
for the QCD sum rule approach (pole model), respectively.

\subsection{Polarization Asymmetries}

We now discuss the longitudinal, normal and transverse polarization
asymmetries of the lepton and their implications. From Eqs. (\ref{pl})$-$(%
\ref{pt}), the distributions of $P_{L}$, $P_{N}$ and $P_{T}$ with respect to
the dimensionless kinematic variable $\hat{s}$ are shown in Figures $4-9$,
respectively. From the figures, we find that the results of the QCD sum rule
and pole models to various polarizations are as follows: (1) they overlap
fully for $P_{L}$; (2) $P_{N}$ is not sensitive to the models except for the
small $q^{2}$ region in $\Lambda_{b}\rightarrow \Lambda \mu ^{+}\mu ^{-}$;
and (3) the effects of the different QCD models to $P_{T}$ are significant
at the large $q^{2}$ region. Clearly, $P_L$ and $P_N$ for the most $q^2$
region in $\Lambda_b\to\Lambda l^+l^-$ are independent of the QCD models.

It is easily seen that outside the resonant states, both polarizations of $%
P_{L}$ and $P_{N}$ are insensitive to the LD effects. We note that $P_L$ for
$\Lambda_b\to\Lambda\mu^+\mu^-$ is close to $1$, while that for the tau mode
is over 40\%, in the most values of $q^{2}$ except that around resonant
regions. The large asymmetries in $\Lambda_b\rightarrow \Lambda l^{+}l^{-}$
are good candidates to test the standard model. For $P_{T}$, since it is
proportional to the imaginary parts of the Wilson coefficient products, the
LD contributions are important. Note that in the standard model, the
effective Wilson coefficients of $C_{9}^{eff}$ contains absorptive parts,
while $C_{7}$ and $C_{10}$ have only real values. From Eq. (\ref{pt}), the
part of $Im(C_{9}^{eff}C_{10}^{*})$ yields a nonzero value of $P_{T}$, but
that of $Im(C_{7}C_{10}^{*})$ vanishes. However, due to the enhanced factor $%
1/\hat{s}$ at small $\hat{s}$ for the term corresponding to $%
Im(C_{7}C_{10}^{*})$, one could search for these regions since the
contribution from some non-standard CP violation model may not be negligible.

%For seeing the new physics effects, we choose $C_{7}$ and $C_{9}$ being
%opposite sign to the values of SM to contribute to longitudinal
%polarization asymmetries which have larger values among the lepton
%asymmetries of $\Lambda _{b}\rightarrow \Lambda l^{+}l^{-}$ decays. From
%figures 4 and 5, we clearly see that $P_{L}$ are dependent on sign$\left(
%C_{9}\right)$ strongly.

Finally, in Table 2, we list the integrated lepton polarization asymmetries
in $\Lambda_b\to\Lambda l^+l^-$, defined by
\begin{eqnarray}
\bar{P}_{i}&=&\int_{4\hat{m}_{l}^{2}}^{\hat{s}_{\max }}d\hat{s}\ P_{i}\,.
\end{eqnarray}
In the table, the results are calculated in the standard model without LD
effects.

\begin{table}[h]
\caption{ Integrated lepton polarization asymmetries in the standard model
without LD effects.}
\begin{center}
\begin{tabular}{|l|l|l|l|l|}
\hline
&  &  &  &  \\
Model & Mode & $10^2\bar{P}_L $ & $10^2\bar{P}_N $ & $10^2\bar{P}_T$ \\
\hline
QCD sum rule & $\Lambda _{b}\rightarrow \Lambda \mu ^{+}\mu ^{-}$ &
\multicolumn{1}{|c|}{$58.3$} & \multicolumn{1}{|c|}{$-9.4$} & $-0.07$ \\
\cline{2-5}
& $\Lambda _{b}\rightarrow \Lambda \tau ^{+}\tau ^{-}$ &
\multicolumn{1}{|c|}{$10.9$} & \multicolumn{1}{|c|}{$-10.0$} & $-0.39$ \\
\hline
pole model & $\Lambda _{b}\rightarrow \Lambda \mu ^{+}\mu ^{-}$ &
\multicolumn{1}{|c|}{$58.3$} & \multicolumn{1}{|c|}{$-12.6$} & $-0.07$ \\
\cline{2-5}
& $\Lambda _{b}\rightarrow \Lambda \tau ^{+}\tau ^{-}$ &
\multicolumn{1}{|c|}{$10.9$} & \multicolumn{1}{|c|}{$-9.2$} & $-0.34$ \\
\hline
\end{tabular}
\end{center}
\end{table}

\section{Conclusions}

We have given a detailed analysis on the dilepton forward-backward and the
longitudinal, normal and transverse lepton polarization asymmetries for the
decays of $\Lambda_b\to\Lambda l^+l^-$ ($l=e,\mu,\tau$) in the standard
model. Based on the HQET, there are only two independent form factors, $%
F_{1} $ and $F_{2}$ or $F_{1}$ and $R$, involved in the matrix element of $%
\Lambda_b\to\Lambda$.

We have shown that all the asymmetries are related to $R$ and free of the
other form factor $F_1$. Moreover, we have found that $R$ is always
associated with $\sqrt{r}$ so that by neglecting their contributions, there
are only a few percentages lose in the asymmetries. Thus, the asymmetries in
the heavy baryonic dilepton decays have a less dependence on the
non-perturbative QCD effects . We have also demonstrated that $%
P_L(\Lambda_b\to\Lambda l^+l^-)$ are QCD model independent quantities. We
have pointed out that the FBA for the light lepton mode gets to zero at $%
\hat{s}_0$ which is only sensitive to the weak couplings. Finally, since the
absolute values of the integrated T-odd observables of the transverse lepton
polarizations in $\Lambda_b\to\Lambda l^+l^-$ are less than $10^{-2}$ in the
standard model, measuring $P_T$ such as in the tau model at a level of $%
10^{-2}$ would be a clear signal for some new CP violation.\newline

{\bf Acknowledgments}

This work was supported in part by the National Science Council of the
Republic of China under contract numbers NSC-89-2112-M-007-054 and
NSC-89-2112-M-006-033.

\newpage
\begin{figcap}
\item
% Fig.1:
FBAs as a function of $q^2/M^2_{\Lambda_b}$
 for (a) $\Lambda_b\to\Lambda\mu^+\mu^-$  and
(b) $\Lambda_b\to\Lambda\tau^+\tau^-$.
The curves with and without resonant shapes represent including and no
LD contributions, respectively.
The solid (dash-dotted) curves stand for the QCD sum rule approach and
 the dashed (dotted) for the pole model
with (without) $R$, respectively.

\item
% Fig.2:
The differential FBA of
$dA_{FB}/dq^2$ for $\Lambda_b\to\Lambda\mu^+\mu^-$
as a function of $q^2$.
Legend is the same as Figure 1.

\item
% Fig.3:
Longitudinal polarization asymmetries.
Legend is the same as Figure 1.

\item
% Fig.4:
Normal polarization asymmetries.
Legend is the same as Figure 1.
\item
% Fig.5:
Transverse polarization asymmetries.
Legend is the same as Figure 1.

\end{figcap}

\newpage
\begin{figure}[h]
\includegraphics{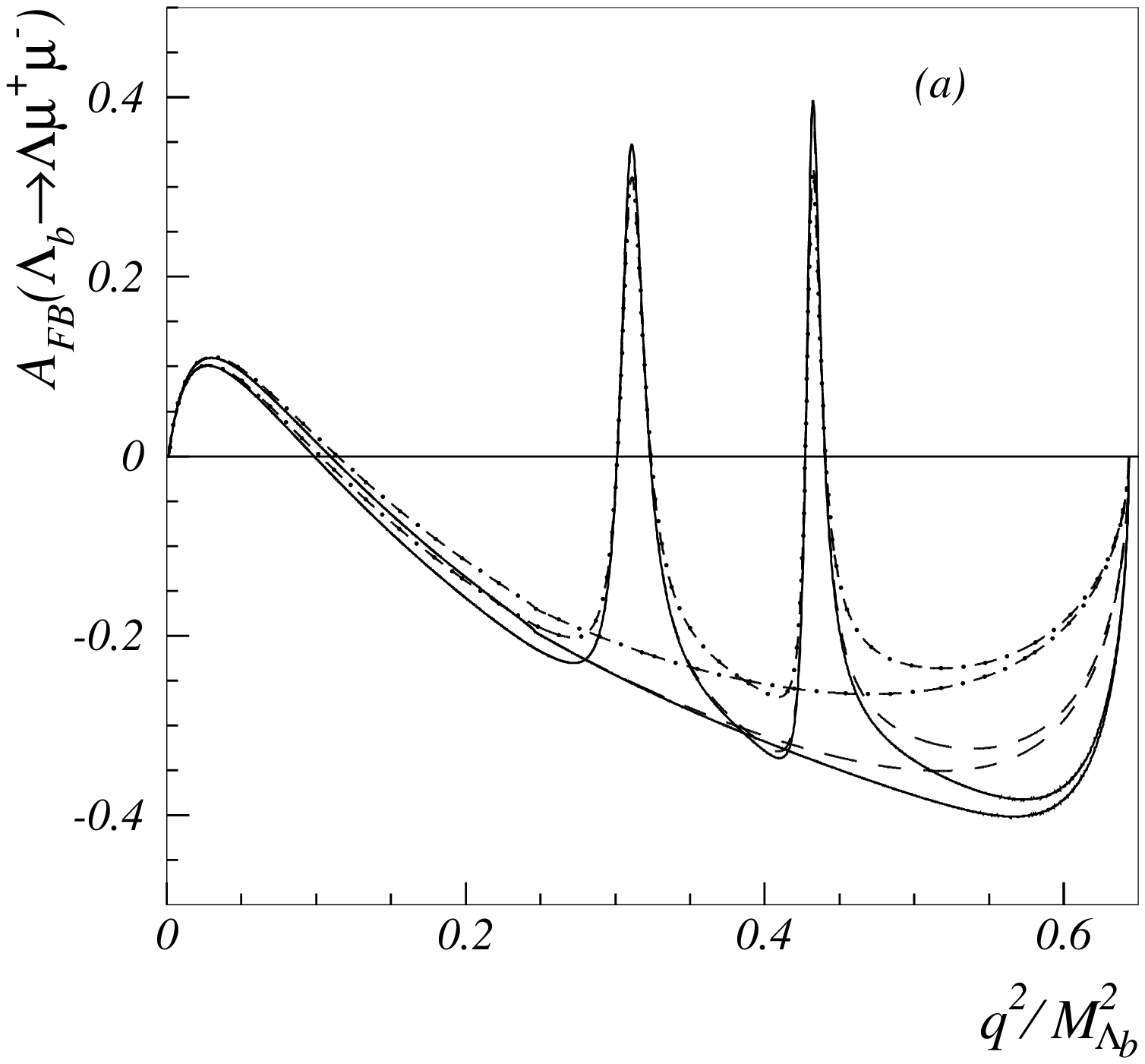}
\vskip 5.5cm
\end{figure}

\vskip 2.cm
\begin{figure}[h]
\includegraphics{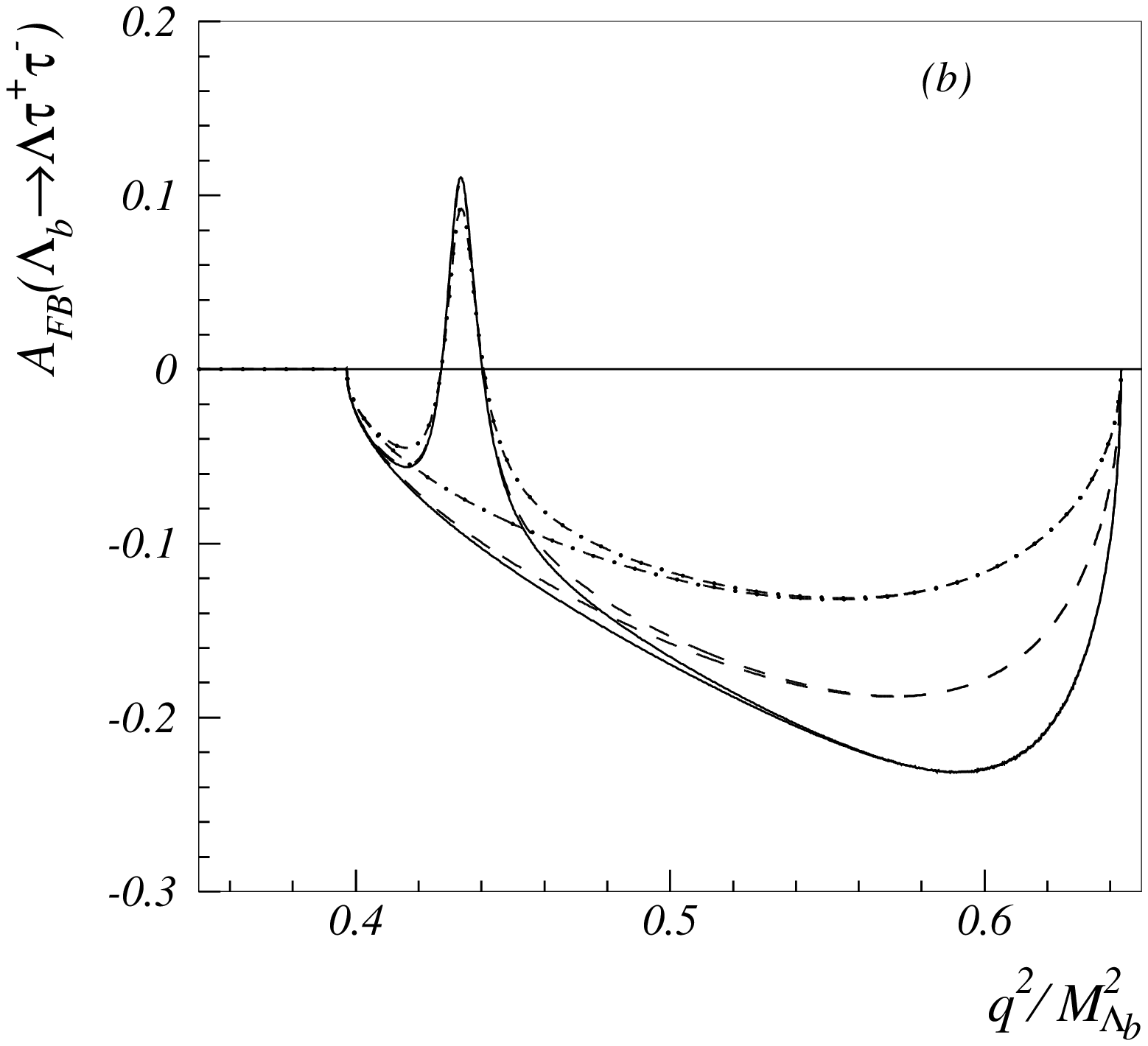}
\vskip 8.cm
\caption{ FBAs as a function of $q^2/M^2_{\Lambda_b}$ for (a) $%
\Lambda_b\to\Lambda\mu^+\mu^-$ and (b) $\Lambda_b\to\Lambda\tau^+\tau^-$.
The curves with and without resonant shapes represent including and no LD
contributions, respectively. The solid (dash-dotted) curves stand for the
QCD sum rule approach and the dashed (dotted) for the pole model with
(without) $R$, respectively. }
\end{figure}

\newpage
\begin{figure}[h]
\includegraphics{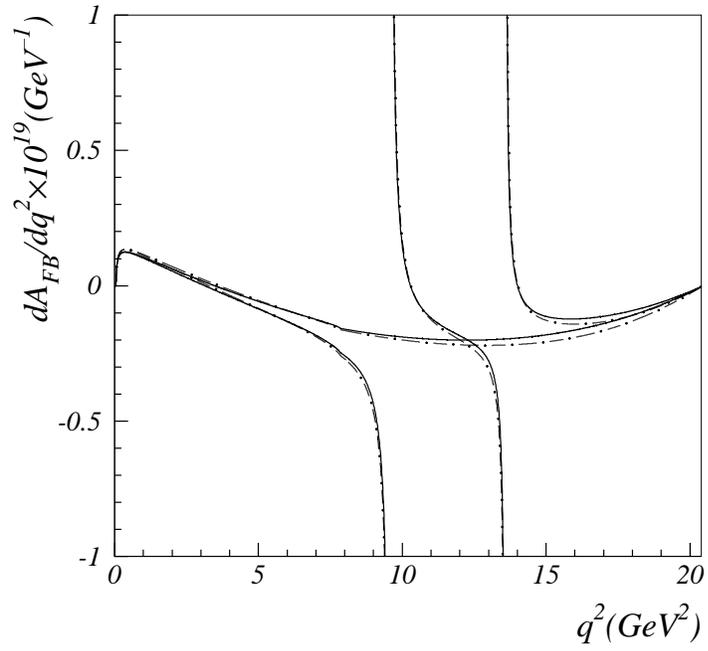} %
\vskip 13cm
\caption{ The differential FBA of $dA_{FB}/dq^2$ for $\Lambda_b\to\Lambda%
\mu^+\mu^-$ as a function of $q^2$. Legend is the same as Figure 1. }
\end{figure}

\newpage
\begin{figure}[h]
\includegraphics{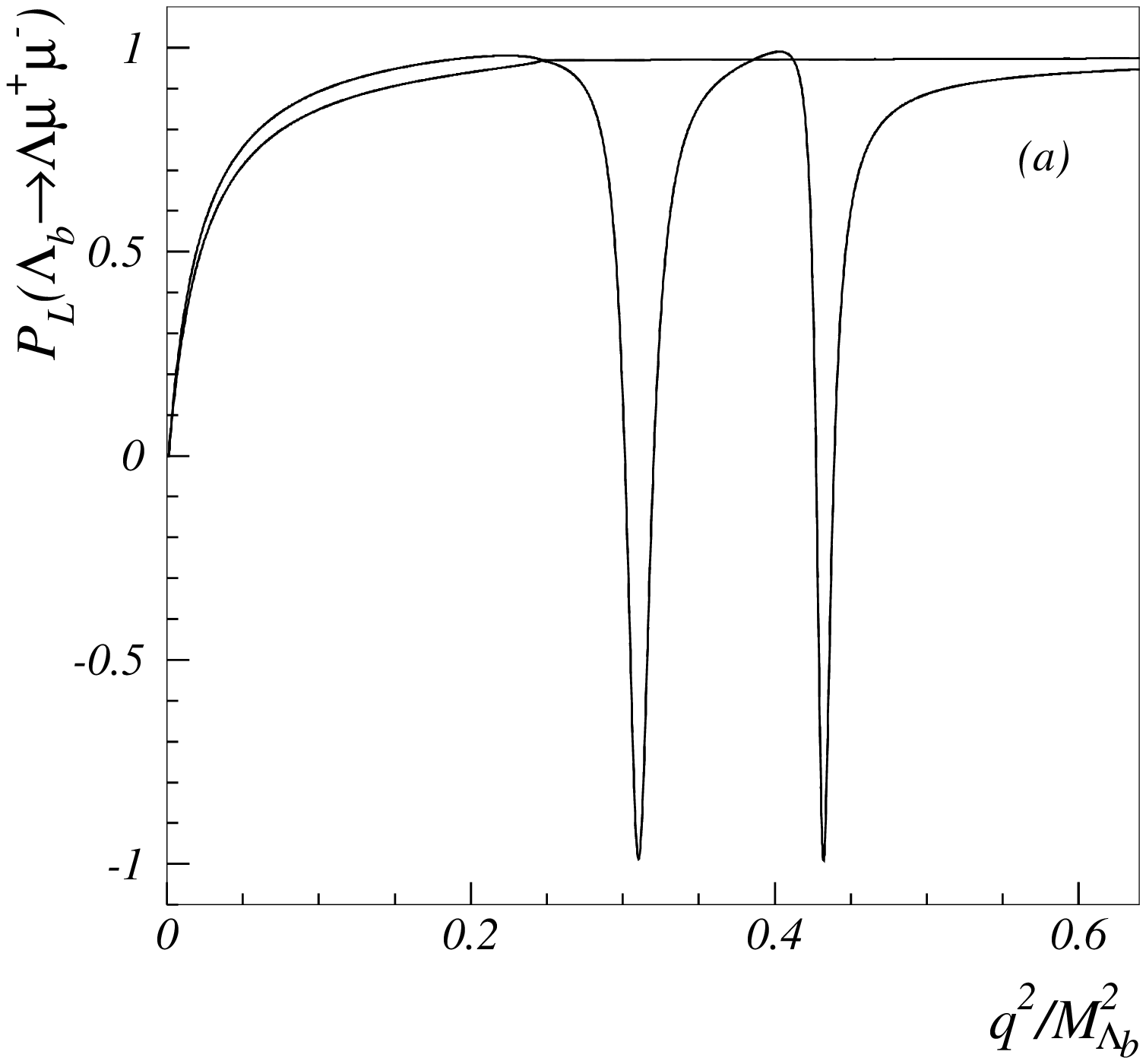}
\vskip 5.5cm
\end{figure}

\vskip 2.cm
\begin{figure}[h]
\includegraphics{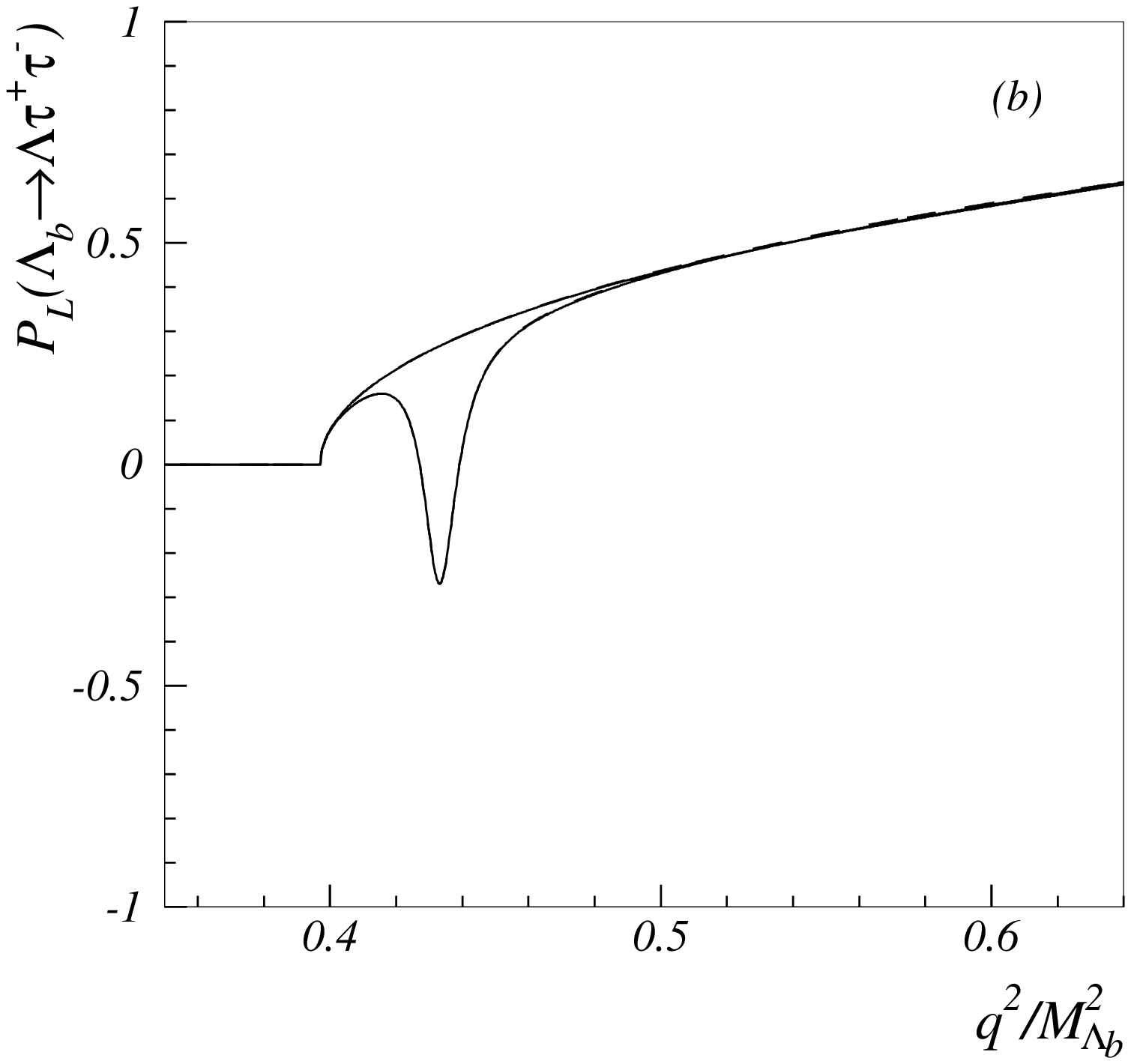}
\vskip 8.cm
\caption{ Longitudinal polarization asymmetries. Legend is the same as
Figure 1. }
\end{figure}

\newpage
\begin{figure}[h]
\includegraphics{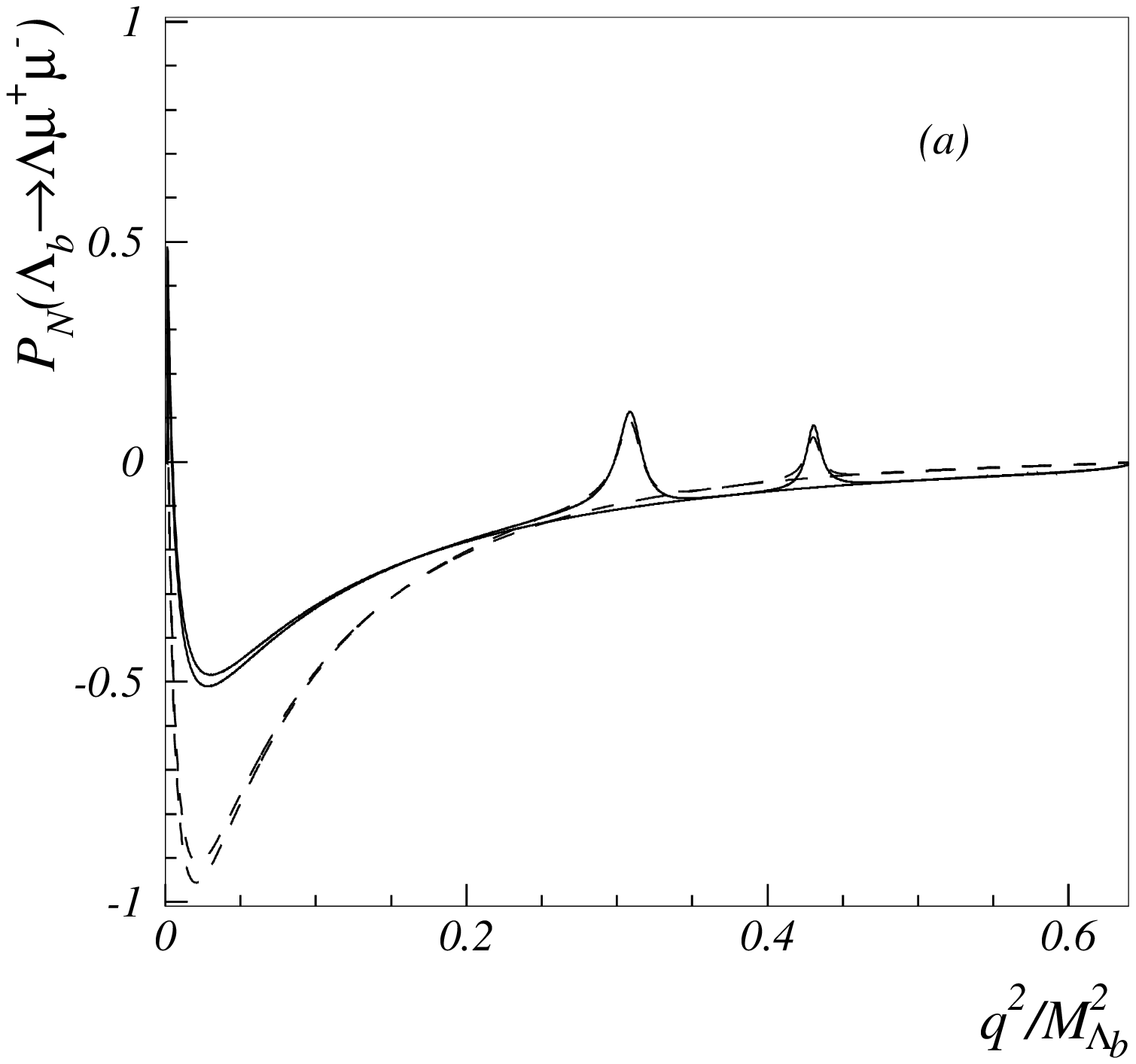}
\vskip 5.5cm
\end{figure}

\vskip 2.cm
\begin{figure}[h]
\includegraphics{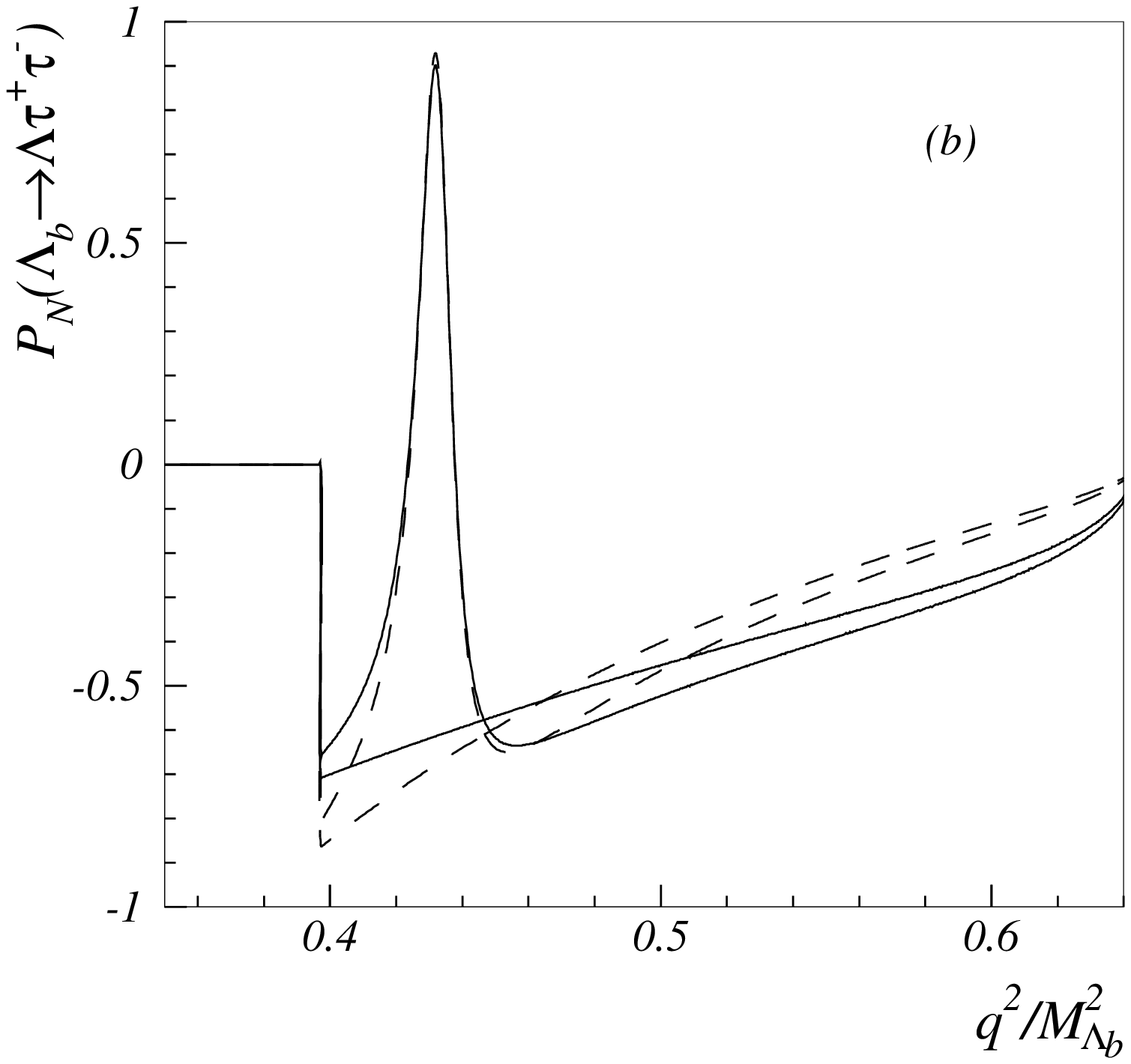}
\vskip 8.cm
\caption{ Normal polarization asymmetries. Legend is the same as Figure 1. }
\end{figure}

\newpage
\begin{figure}[h]
\includegraphics{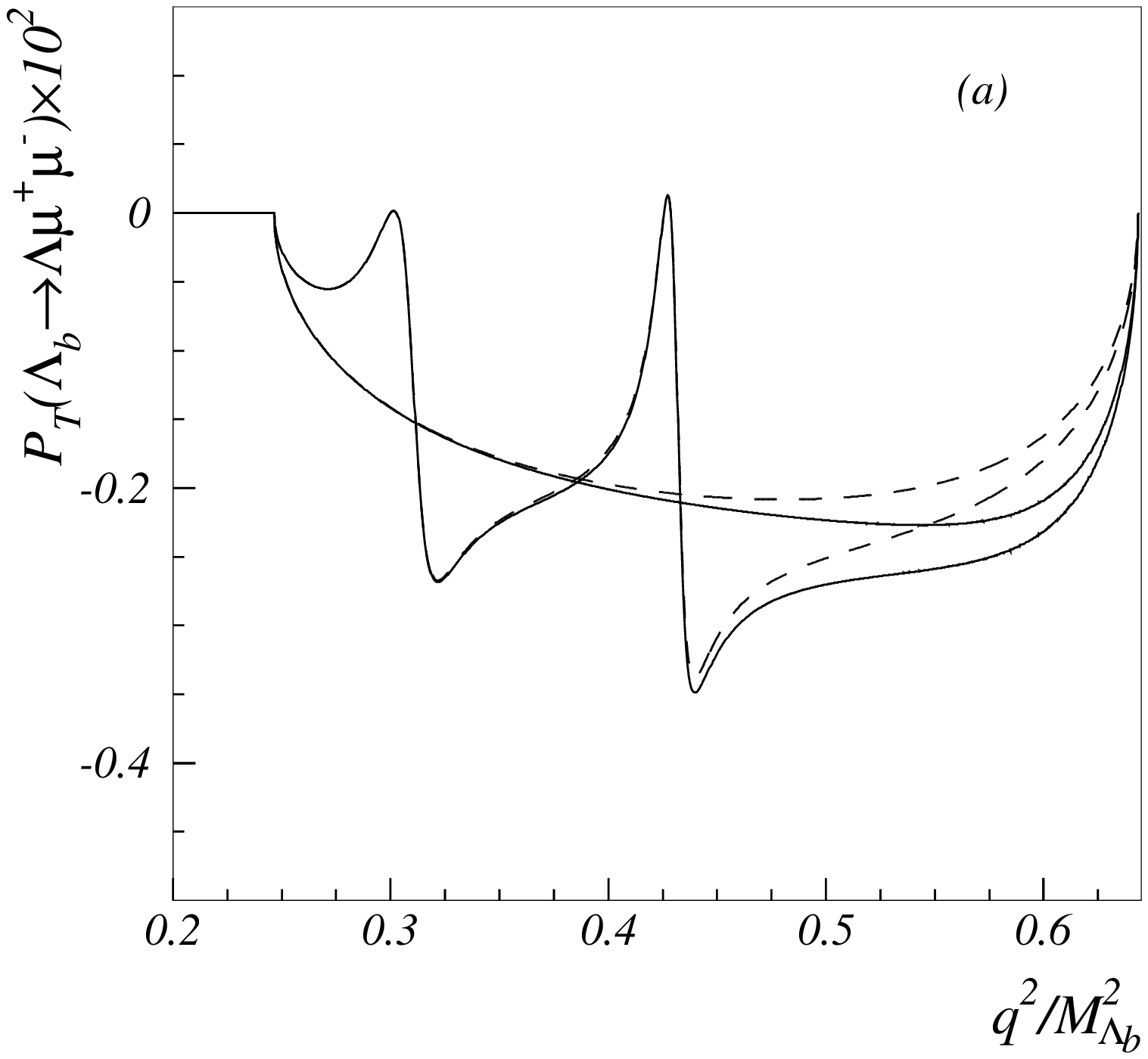}
\vskip 5.5cm
\end{figure}

\vskip 2.cm
\begin{figure}[h]
\includegraphics{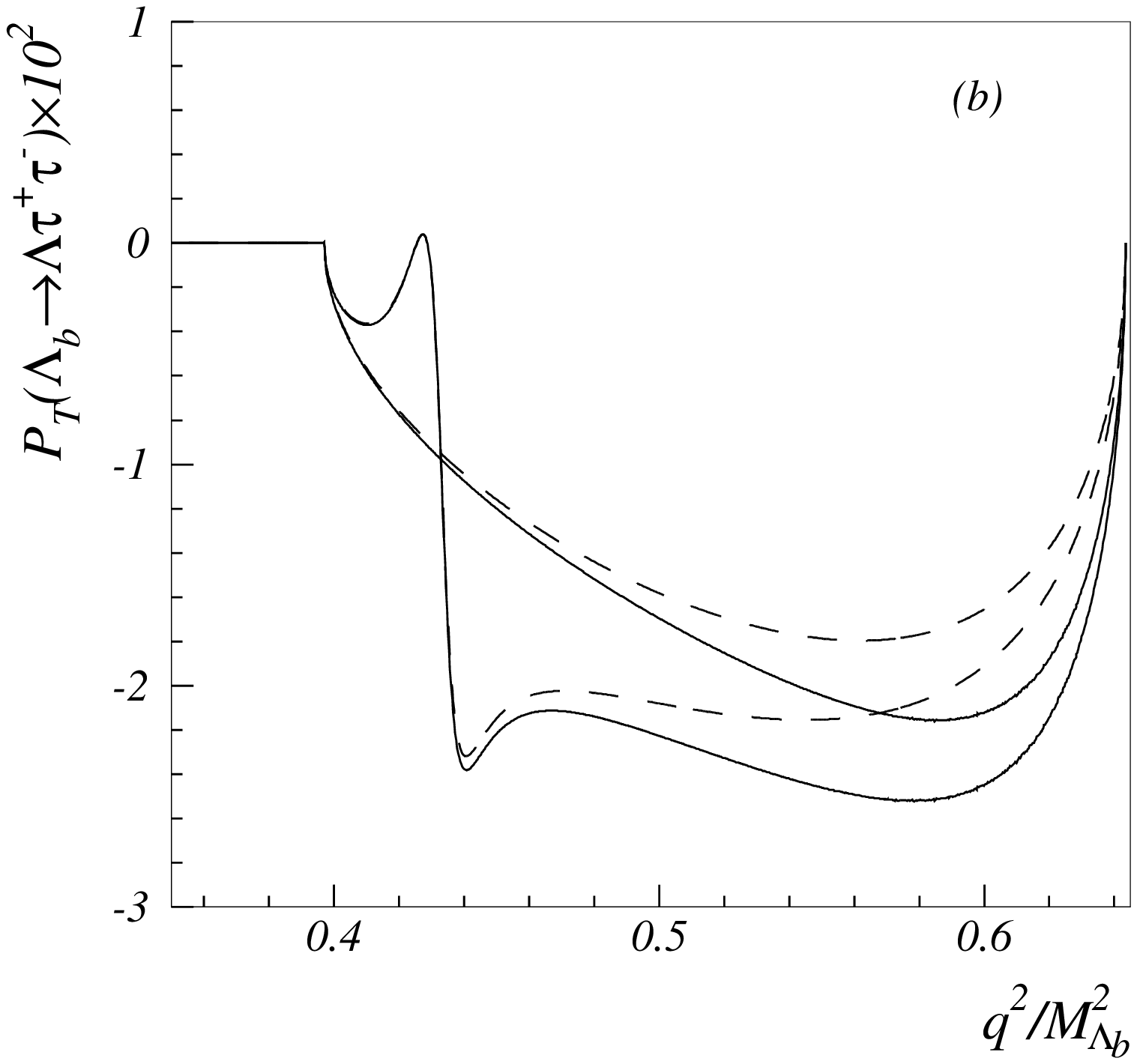}
\vskip 8.cm
\caption{ Transverse polarization asymmetries. Legend is the same as Figure
1. }
\end{figure}

\end{document}